\documentstyle[12pt]{article}

\columnwidth\textwidth

\newcommand{\B}{{\cal B}}
\newcommand{\unit}[1]{\,\rm{#1}}
\newcommand{\BW}{\mbox{BW}}

\begin{document}
\pagenumbering{arabic}
\everymath={\displaystyle}
\vspace*{-2mm}
\thispagestyle{empty}
\noindent
\hfill hep-ph/9706465\\
\mbox{}
\hfill HUTP-97/A028\\
\mbox{}
\hfill \today \\
\begin{center}
  \begin{Large}
  \begin{bf} {\Large \sc 
The Chirally Normalized Vector Meson Dominance Model and
Semihadronic Tau Decays}
   \\
  \end{bf}
  \end{Large}
  \vspace{0.8cm}
   Markus Finkemeier \\[2mm]
   {\em Lyman Laboratory of Physics\\
        Harvard University\\
        Cambridge, MA 02138, USA}\\[5mm]
{\bf Abstract} 
\end{center}
\noindent
We discuss and test the assumptions of the chirally normalized
vector meson dominance model (CN-VMD), as it is used for tau decays.
We compare the CN-VMD amplitudes with those using on-shell
couplings, derived directly from experimental data. We discuss in detail the 
$3\pi$, $K K \pi$ and $2\pi$ final states. We suggest that the 
true values for the $K_1$ widths might be larger than those
quoted in the Review of Particle Physics.
%
%
%
\newpage

\section{Introduction}       

We (R.\ Decker, J.\ K\"uhn, E.\ Mirkes, the present author and others) 
\cite{we,kaon}
have discussed semihadronic tau decays in several papers.
The model we have employed might be called the 'chirally normalized
vector meson dominance' or CN-VMD model, and has been invented and
used by others before \cite{others}. 
It rests on two basic assumptions: (i) The 
decay amplitudes are assumed to be fully dominated by intermediate
resonant states (VMD). (ii) The meson couplings are assumed to be
constant from $Q^2 = 0$ up to the resonance masses. Matching the VMD 
model to the chiral limit then allows one to fix the relevant products
of meson couplings without additional experimental input ('chiral 
normalization', CN).

In addition to these two basic assumptions of the CN-VMD model, 
additional assumptions have to be made: Which resonances are included:
Only the lowest lying ones, or also higher radials, and if yes, with
which relative contribution? Only vector resonances ($\rho$, $K^*$, 
\dots), or also axial vector resonances? Furthermore, we normally 
assume exact $SU(3)$ flavor symmetry for the couplings, even if we
use the physical masses $m_\pi$, $m_K$, \dots and widths for the 
mesons.

Alternatives to the CN-VMD model have been discussed by Oakes, Braaten
{\em et al.} \cite{braaten}, and by Li \cite{li}. 
The purpose of the present note is to
test some of the assumptions of the CN-VMD model such as the
constancy of the couplings, and to give some indications how the model
could be refined.
\section{The Chiral Normalization for $\tau \to 3\pi \nu_\tau$}       

In this section we will try to test the assumption of constancy of
the meson couplings for the three pion decay mode. 
In order to do that, we compare the CN-VMD model with a VMD model 
where we take the meson couplings $g_{\rho\pi\pi}$ and 
$g_{a_1 \rho \pi}$ directly from experimental data \cite{braaten}.
This VMD is
also based on the decay chain
$$
   \tau \to a_1 \nu_\tau, \quad a_1 \to \rho \pi, \quad \rho \to \pi \pi
$$
Essentially, the idea is that meson couplings taken from experimental data
are measured at $Q^2$ of $m_\rho^2$ or $m_{a_1}^2$ (on-shell couplings). 
If the assumption is
right that the meson couplings are (approximately) constant from these
$Q^2$ of the order of $1 \unit{GeV}^2$ down to $Q^2 = 0$, then this model
and the CN-VMD model should give the same amplitudes.

One immediate problem is that we don't know the coupling $f_{a_1}$
of the $a_1$ to the weak axial current very well. We will instead
rely on perturbative QCD, pQCD \cite{pQCD,suzuki}.
From pQCD, we know that
\begin{equation}
   (30.1 \pm 0.1)\% = \B(\pi) + \B(a_1)
\end{equation}
where we assume vector meson dominance (VMD) of the $J^P = 1^+$
spectral function\footnote{
Note that this assumption on a branching ratio level is much weaker
that the VMD assumption on the amplitude level in the CN-VMD model.
If there is an additional non-resonant contribution,
say a direct transition $W \to 3 \pi$, it will have a very small 
contribution to the total branching ratio into the axial vector channel,
even if at very small $Q^2$, its amplitude would be of the same order
than that of the $W \to a_1 \to 3 \pi$ channel.
}
(in the sense that only the $\pi$ and the $a_1$ resonant intermediate
state contribute to this spectral function).
With our VMD assumption, $\B(a_1) \geq \B(3\pi) + B(K K \pi)_A$.
The subscript 'A' denotes the axial vector part.

With the conventions given in the appendix, we have
\begin{equation}
   \Gamma(\tau \to a_1 \nu_\tau) =
   \frac{m_\tau^3}{8 \pi} \left[ \frac{G_F f_{a_1} \cos \theta_c}
   {m_{a_1}} \right]^2 \left[1 - \frac{m_{a_1}^2}{m_\tau^2} \right]^2
   \left[1 + \frac{2 m_{a_1}^2}{m_\tau^2} \right]
\end{equation}
and thus
\begin{equation}
  f_{a_1} = 0.2050 \unit{GeV}^2
\end{equation}

Next we will relate $g_{a_1 \rho \pi}$ to $\Gamma_{a_1}$ and
$g_{\rho\pi\pi}$ to $\Gamma_\rho$. Consider the generic decay of
a spin-1 particle with mass $M$ into two spin-0 particles with
masses $m_1$ and $m_2$. We assume the matrix element is given by
\begin{equation}
   {\cal M} = g_{100} \epsilon \cdot (p_1 - p_2)
\end{equation}
(where $\epsilon$ is the polarization vector of the spin-1 particle and
$p_i$ are the momenta of the spin-0 particles). Then
$$
   \Gamma[(J=1) \to (J=0) + (J=0]] 
$$
\begin{equation}
   = 
   \frac{g_{100}^2}{48 \pi} \frac{\sqrt{[M^2 - (m_1+m_2)^2]
   [M^2 - (m_1 - m2)^2]}}{M} 
   \left[1 - 2 \frac{m_1^2 + m_2^2}{M^2} + \left(\frac{m_1^2 - m_2^2}
   {M^2}\right)^2 \right]
\end{equation}
Thus from $\Gamma(\rho\to\pi\pi) = 150.7\unit{MeV}$, we find that
\begin{equation}
   g_{\rho\pi\pi} = 6.049
\end{equation}
(if we neglected the pion mass by putting $m_\pi = 0$, we would get
$g_{\rho\pi\pi} = 5.438$). 

Similarly, for a generic decay of a spin-1 particle with mass $M$ 
into a spin-1 particle, mass $m_1$ and a spin-0 particle, mass $m_2$,
we assume the matrix element is given by
\begin{equation}
   {\cal M} = i g_{110} \epsilon^\mu(p) \epsilon_mu(p_1)
\end{equation}
where $p$ and $p_1$ are the momenta of the initial and final spin-1 particles,
respectively. Note that we are ignoring four possible additional form
factors here. The contribution of three of these vanish for on-shell
particles, and there is one which is of higher order in the momenta,
but which could contribute on-shell. The above ansatz leads to
$$
   \Gamma[(J=1) \to (J=1) + (J=0)] 
$$
\begin{equation}
   = \frac{g_{110}^2}{192\pi} \frac{\sqrt{[M^2-(m_1+m_2)^2]
   [M^2-(m_1-m_2)^2]}}{M^3} 
   \left[8 + \frac{(M^2 - m_2^2 + m_1^2)^2}{M^2 m_1^2} \right]
\end{equation}

Neglecting the pion mass, we find from this
\begin{equation}
   \Gamma(a_1 \to \rho \pi) = \frac{1}{96 \pi} \frac{g_{a_1 \rho\pi}^2}
   {m_{a_1}} \left[1 - \frac{m_\rho^2}{m_{a_1}} \right]
   \left[10 + \frac{m_{a_1}^2}{m_\rho^2} + \frac{m_\rho^2}{m_{a_1}^2} 
\  \right]
\end{equation}
If we assume $\Gamma_{a_1} = 400 \unit{MeV}$, this yields
\begin{equation}
   g_{a_1 \rho \pi} = 4.33 \unit{GeV}
\end{equation}

Plugging all together, we get for the amplitude 
\begin{equation}
   {\cal M} (\tau\to 3\pi) = \frac{G_F}{\sqrt{2}} \cos \theta_C
   [\overline{u}_\nu \gamma_\mu \gamma_- u_\tau] H^\mu
\end{equation}

In the VMD model with experimentally determined on-shell couplings,
we have
\begin{equation}
   H^\mu = - i \frac{\sqrt{2} f_{a_1} g_{a_1 \rho \pi} g_{\rho\pi\pi}}
   {m_{a_1}^2 m_\rho^2} \BW_{a_1}(Q^2) \BW_\rho(s_1)
   (p_2 - p_3)_\nu T^{\mu\nu} + (1 \leftrightarrow 2)
\end{equation}
where $\BW_X(s)$ are the usual normalized resonance factors with
$\BW_X(0) = 1$, and 
$$
   T^{\mu\nu} = g^{\mu\nu} - \frac{Q^\mu Q^\nu}{Q^2}
$$

Note that these $f_{a_1}$, $g_{a_1 \rho \pi}$ and $g_{\rho \pi\pi}$ 
are the on-shell values (all mesons on their various mass shells). 
Of
course, the concept of on-shell resonances 
is an approximation for the $a_1$ and, to a lesser extent,
for the $\rho$, because of their non negligible widths.

The chiral limit of the hadronic current is
\begin{equation}
   H^\mu = -i \frac{2\sqrt{2}}{3 f_\pi} \left\{ (p_2 - p_3) T^{\mu\nu}
   + (1 \leftrightarrow 2) \right\}
\end{equation}
This implies the following relation for the couplings {\em at zero
momenta}, which we will denote by $\overline{f}_{a_1}$, 
$\overline{g}_{a_1\rho\pi}$ and $\overline{g}_{\rho\pi\pi}$:
\begin{equation}
   \overline{f}_{a_1} \overline{g}_{a_1\rho\pi} \overline{g}_{\rho\pi\pi}
   = \frac{2 m_{a_1}^2 m_\rho^2}{3 f_\pi} = 6.461 \unit{GeV}^3
\end{equation}
where I used $f_\pi = 92.1\unit{MeV}$ \cite{fpi}.

The product of the on-shell values is
\begin{equation}
   {f}_{a_1} {g}_{a_1\rho\pi} {g}_{\rho\pi\pi} =
   4.827 \unit{GeV}^3
\end{equation}

So we find
\begin{equation}
   \frac{{f}_{a_1} {g}_{a_1\rho\pi} {g}_{\rho\pi\pi}}
   {\overline{f}_{a_1} \overline{g}_{a_1\rho\pi} \overline{g}_{\rho\pi\pi}}
   = \left\{ \begin{array}{l@{\quad}l}
   0.831 & \mbox{for $\Gamma_{a_1} = 400 \unit{MeV}$} \\
   0.881 & \mbox{for $\Gamma_{a_1} = 450 \unit{MeV}$} \\
   1.018 & \mbox{for $\Gamma_{a_1} = 600 \unit{MeV}$} 
   \end{array} \right.
\end{equation}
where we have indicated the dependence on the value of $\Gamma_{a_1}$
chosen. We find that the ratio of the value of the product of the
couplings at on-shell momenta divided by their product at $Q^2=0$ 
is reasonably close to one. However, the uncertainty in the $a_1$ width
does not allow a very precise conclusion, and indeed the large value
of $\Gamma_{a_1}^2 / m_{a_1}^2$ makes the whole notion of on-shell
couplings ill-defined. 

If we want to, we can fit $\Gamma_{a_1}$ to make the ratio equal to
one, resulting in $\Gamma_{a_1} = 579 \unit{MeV}$. For the reasons
given, this value should, however, not be taken too seriously.

\section{The $K_1$ and $\tau \to K \pi \pi \nu_\tau$}

The situation is complicated by $SU(3)_F$ flavor symmetry breaking and 
the $K_1^A - K_1^B$ mixing. For simplicity, we will first assume
\begin{equation}
   m_{K_1(1270)} \approx m_{K_1(1400)} = 1340 \unit{MeV}
\end{equation}
and neglect the existence of two $K_1$ resonances.

With this assumption, using the pQCD prediction for the
strange axial / pseudoscalar hadronic current, we have
\begin{equation}
   \B(K_1) = (1.31 \pm 0.06)\% - \B(K) = (0.59 \pm 0.06)\%
\end{equation}
This results in
\begin{equation}
   f_{K_1} = (0.2026 \pm 0.0103) \mbox{GeV}
\end{equation}

Note that
\begin{equation} \label{eqn1}
   \frac{f_{K_1}}{f_{a_1}} = 0.9883
\end{equation}
and
\begin{equation} \label{eqn2}
   \frac{f_{K_1}/m_{K_1}^2}{f_{a_1}/m_{a_1}^2} = 0.8372
\end{equation}
It appears that pQCD prefers $SU(3)$ symmetry in the simpler version
(\ref{eqn1}) over (\ref{eqn2}).

Next
\begin{equation}
   \Gamma(K_1^- \to K^{*0} \pi^-) = 
   \frac{1}{192 \pi} \frac{g_{K_1 K^* \pi}^2}{m_{K_1}}
   \left[1 - \frac{m_{K^*}^2}{m_{K_1}^2} \right]
   \left[10 + \frac{m_{K_1}^2}{m_{K^*}^2} 
   + \frac{m_{K^*}^2}{m_{K_1}^2} \right]
\end{equation}
Using
\begin{equation}
   \Gamma(K_1^- \to K^{*0} \pi^-) = 
   \frac{2}{3} \times 94\% \times 174 \unit{MeV} = 109 \unit{MeV}
\end{equation}
we obtain
\begin{equation}
   g_{K_1 K^* \pi} = 3.535 \unit{GeV}
\end{equation}

Along similar lines, 
$$
   \Gamma(K_1^- \to \rho^0 K^-) =
   \frac{g_{K_1 \rho K}^2}{192\pi} 
   \frac{\sqrt{[m_{K_1}^2 - (m_\rho + m_K)^2]
   [m_{K_1}^2 - (m_\rho - m_K)^2]}}{m_{K_1}^3}
$$
\begin{equation} \times
   \left[8 + \frac{(m_{K_1}^2 - m_K^2 + m_\rho^2)^2}
   {m_{K_1}^2 m_\rho^2} \right]
\end{equation}
With $\Gamma(K_1^- \to \rho^0 K^-) = 12.6\unit{MeV}$, we find
\begin{equation}
   g_{K_1 \rho \pi} = 1.5848 \unit{GeV}
\end{equation}

Finally, from $\Gamma(K^* \to K \pi)$, 
\begin{equation}
   g_{K^* K \pi} = 4.578
\end{equation}

Now consider the decay
$$
  \tau \to K^- \pi^- \pi^+
$$
Plugging everything together, we find the hadronic current $H^\mu$
in a model with experimentally determined on-shell couplings:
$$
  H^\mu = \frac{\sqrt{2} f_{K_1}}{m_{K_1}^2}
  \BW_{K_1}(Q^2) T^{\mu\nu} \left\{
  \frac{g_{K_1 K^* \pi} g_{K^* K \pi}}{m_{K^*}^2} \BW_{K^*}(s_2)
  (p_1 - p_3)_\nu \right.
$$
\begin{equation}
  + \left.
  \frac{g_{K_1 \rho K} g_{\rho\pi\pi}}{m_{\rho}^2} \BW_{\rho}(s_1)
  (p_2 - p_3)_\nu \right\}
\end{equation}
The same current in the CN-VMD, on the other hand, is
\begin{equation}
  H^\mu = \frac{\sqrt{2}}{3 f_\pi}
  \BW_{K_1}(Q^2) T^{\mu\nu} \left\{
   \BW_{K^*}(s_2)
  (p_1 - p_3)_\nu
  +
   \BW_{\rho}(s_1)
  (p_2 - p_3)_\nu \right\}
\end{equation}
Equating the current with on-shell couplings and the CN-VMD 
results in two equations
\begin{equation} \label{eqni}
   \frac{\sqrt{2}}{3 f_\pi} \stackrel{!}{=}
   \frac{\sqrt{2} f_{K_1} g_{K_1 K^* \pi} g_{K^* K \pi}}
   {m_{K_1}^2 m_{K^*}^2}
\end{equation}
and
\begin{equation} \label{eqnii}
   \frac{\sqrt{2}}{3 f_\pi} \stackrel{!}{=}
   \frac{\sqrt{2} f_{K_1} g_{K_1 \rho K} g_{\rho\pi\pi}}
   {m_{K_1}^2 m_{\rho}^2}
\end{equation}
Numerically, the left hand side of both equations is
$$
   \frac{\sqrt{2}}{3 f_\pi}  = 5.118 \unit{GeV}^{-1}
$$
The right hand side of (\ref{eqni}) is
$$
   \frac{\sqrt{2} f_{K_1} g_{K_1 K^* \pi} g_{K^* K \pi}}
   {m_{K_1}^2 m_{K^*}^2} = 3.245 \unit{GeV}^{-1} 
$$
and the RHS of (\ref{eqnii}) is
$$
   \frac{\sqrt{2} f_{K_1} g_{K_1 \rho K} g_{\rho\pi\pi}}
   {m_{K_1}^2 m_{K^*}^2} = 2.593 \unit{GeV}^{-1} 
$$

So we find that both equations do not work well. This can
be explained by energy dependence of the meson couplings.
This interpretation implies
\begin{eqnarray}
   \frac{f_{K_1} g_{K_1 K^\star \pi} g_{K^\star K \pi}}
   {\overline{f}_{K_1} \overline{g}_{K_1 K^\star \pi} 
    \overline{g}_{K^\star K \pi}} & = & 0.634
\\ \nonumber 
\\ \nonumber 
   \frac{f_{K_1} g_{K_1 \rho K} g_{\rho\pi \pi}}
   {\overline{f}_{K_1} \overline{g}_{K_1 \rho \pi} 
    \overline{g}_{\rho\pi\pi}} & = & 0.5067
\end{eqnarray}
As before, couplings without bar are on-shell, and with bar they
denote their values at $Q^2 = 0$.
This would imply that the assumption of constancy of the coupling,
which is made in the CN-VMD model, is violated by a large amount.

Note, however, that the size of the violation of the constancy 
appears very large, when compared with the $a_1 \to 3\pi$ case,
where this assumption seemed to work rather well.
This observation appears to remain true, in spite of the fact that $SU(3)$ 
is an additional source for violation of (\ref{eqni}, \ref{eqnii}).
A stronger statement here would require a more sophisticated study
of $SU(3)$ flavor symmetry breaking.

However, note that
a crucial input in evaluating the RHS's in (\ref{eqni}, \ref{eqnii})
are the $K_1$ widths.
As of now, these have only been measured in hadronically production.
But remember the case of the $a_1$. Hadronic production yielded rather
small values for $\Gamma_{a_1}$, about $(250 \cdots 300) \unit{MeV}$.
Measurements of $\Gamma_{a_1}$ gave much larger values. As is
explained in a mini-review in the Review of Particle Physics
\cite{rpp},
the value extracted for the width depends on the assumption for the
form of the coherent background amplitude (Deck-amplitude). Changing
the assumptions allowed to reconcile hadronic and tau decay measurements.

But the very same Deck amplitudes have been used in the measurement
of $\Gamma_{K_1}$, as can be seen from the original papers quoted in the
RPP-96. This leads to the 
possibility that the measured
values for the $K_1$ widths, as quoted in the Review of Particle
Physics,
may actually be much too small. 
But we can satisfy (\ref{eqni}, \ref{eqnii}) 
by assuming that
\begin{eqnarray}
   \frac{g_{K_1 K^* \pi}^{\rm{true}}}{g_{K_1 K^* \pi}^{\rm{RPP}}}
   & = & 1.577
\nonumber \\
\nonumber \\
   \frac{g_{K_1 \rho K}^{\rm{true}}}{g_{K_1 \rho K}^{\rm{RPP}}}
   & = & 1.974
\end{eqnarray}

These relations could be satisfied by
\begin{eqnarray}
  \Gamma_{K_1(1270)}^{\rm true} & = & 1.974^2 \Gamma_{K_1(1270)}^{\rm RPP}
  = 350\unit{MeV}
\nonumber \\
\nonumber \\
   \Gamma_{K_1(1400)}^{\rm true} & = & 1.577^2 \Gamma_{K_1(1270)}^{\rm RPP}
  = 433 \unit{MeV}
\end{eqnarray}

Of course, these numbers should be considers as speculative guesses.
Instead, we urge to measure the $K_1$ widths in tau decays, to
settle this matter experimentally.
Also, a re-analysis of the old hadronic $K_1$ widths would be interesting,
with respect to the sensitivity of the results to the precise form
of the assumptions for the coherent background.

\section{The Non-Strange Vector Channel}

From pQCD, and assuming the vector channel to be saturated by
$\rho$ and $\rho'$, we obtain
\begin{equation}
   \B(\tau \to \nu_\tau + (1^-, S=0)) = (31.9 \pm 0.1) \%
   = \B(\rho) + \B(\rho')
\end{equation}

With our conventions, we have
\begin{equation}
   \frac{\B(\rho)}{\B(e)} = 
   \frac{24 \pi^2}{m_\tau^2 m_\rho^2} f_\rho^2 \cos \theta_C^2
   \left[1 - \frac{m_\rho^2}{m_\tau^2} \right]^2
   \left[1 + \frac{2 m_\rho^2}{m_\tau^2} \right]
\end{equation}
and an identical relation for $\rho'$. 

Now $f_{\rho'} / f_\rho$ is {\em known} from $e^+ e^- \to \rho, \,
  \rho' \to \pi^+ \pi^-$ fits to be
\begin{equation}
  \frac{f_\rho'^2}{f_\rho^2} = 2.4 \cdots 2.66 = 2.5 \pm 0.2
\end{equation}
The value $2.4$ is from Suzuki \cite{suzuki}, and the $2.66$ is
from \cite{taurad}. The final number and error is an estimated guess.

Using this to constrain the relative contributions of the $\rho$ and 
the $\rho'$ in the vector channel, we find
\begin{eqnarray}
   f_\rho & = &(0.1167 \pm 0.0025) \mbox{GeV}^2
\nonumber \\
   f_{\rho'} & = &(0.184 \pm 0.007) \mbox{GeV}^2
\end{eqnarray}
where the errors are dominated by $\Delta m_{\rho'}$.

It is interesting to check Weinberg's sum rule with the values 
we have derived.
Following Suzuki \cite{suzuki}, we assume saturation of the sum rules
in the vector channel by the $\rho$ and the $\rho'$, and in the
axial vector-pseudoscalar channel by the $\pi$ and the $a_1$. Then
Weinberg' first sum rule is
\begin{equation}
   \frac{f_{\rho}^2}{m_{\rho}^2} + \frac{f_{\rho'}^2}{m_{\rho'}^2}
   = \frac{f_{a_1}^2}{m_{a_1}^2} + f_\pi^2
\end{equation}
Numerically, this is
$$
  \rm{LHS} = (108.1 \pm 0.3) \% RHS
$$
Note that without the $\rho'$, we would have
$$
  \rm{LHS} = 64 \% RHS
$$
This strongly supports Suzuki's suggestion about which resonances 
dominate the sum rules.

Now the determination of $f_{\rho'}$ is certainly not very precise.
So instead, we could use the Weinberg sum rules to {\em determine}
$f_{\rho'}$. From the first sum rule, we then find
\begin{equation}
   f_{\rho'} = 0.1665 \unit{GeV}^2
\end{equation}

Weinberg's second sum rule becomes
\begin{equation}
   f_\rho^2 + f_{\rho'}^2 = f_{a_1}^2
\end{equation}
which yields
\begin{equation}
   f_{\rho'} = 0.1685 \unit{GeV}^2
\end{equation}
This value is very close to the one derived from the first sum
rule, giving the determination of $f_{\rho'}$ credibility.

Thus we have 
\begin{equation}
   f_{\rho'}  =  \left\{
   \begin{array}{ll}
   (0.184 \pm 0.007) \unit{GeV}^2 & \mbox{from $e^+ e^- \to 2 \pi$} \\
   (0.1675 \pm 0.0010) \unit{GeV}^2 & \mbox{from Weinberg's sum rules}
   \end{array} \right.	
\end{equation}
The two values are about two standard deviations apart, suggesting 
to use 
\begin{equation}
   f_{\rho'}  =  
   (0.1758 \pm 0.0082) \unit{GeV}^2 
\end{equation}
as an average.

Now we can discuss
$$
   \tau \to \pi\pi \nu_\tau
$$

Using a parameterization with experimentally determined on-shell
couplings, we find the relevant hadronic current is
\begin{equation}
  H^\mu = \frac{\sqrt{2} f_\rho g_{\rho\pi\pi}}{m_\rho^2} 
  \BW_\rho(Q^2) (p_1 - p_2)^\mu + (\rho \to \rho')
\end{equation}
Here 
\begin{equation}
   | g_{\rho'\pi\pi} | = 0.23 | g_{\rho\pi\pi} |
\end{equation}
from the experimental data on $\Gamma_{\rho'\to\pi\pi}$. 

{\em If} it is true that the meson couplings are constant down from
on-shell values for the momentum transfers to $Q^2=0$, then we can 
calculate the limit $Q^2 \to 0$ of the above hadronic current:
\begin{equation}
   H^\mu \stackrel{Q^2\to 0}{\to} 
  \left[\frac{\sqrt{2} f_{\rho} g_{\rho\pi\pi}}{m_{\rho}^2} +
    \frac{\sqrt{2} f_{\rho'} g_{\rho'\pi\pi}}{m_{\rho'}^2}
   \right] (p_1 - p_2)^\mu
\end{equation}
The chiral limit predicts
\begin{equation}
   H^\mu \stackrel{Q^2\to 0}{\to} \sqrt{2} (p_1 - p_2)^\mu
\end{equation}

So we can now check the chiral normalization of the VMD model. If it works,
we need 
\begin{equation}
   1 \stackrel{!}{=} \underbrace{
   \underbrace{\frac{f_{\rho} g_{\rho\pi\pi}}{m_{\rho}^2}}_{\displaystyle
   1.197} \pm 
   \underbrace{\frac{f_{\rho'} g_{\rho'\pi\pi}}{m_{\rho'}^2}}_{\displaystyle
   0.116 \pm 0.006}}_{\displaystyle 1.081 \pm 0.006}
\end{equation}
The $\pm$ between the $\rho$ and the $\rho'$ is their relative phase,
which is known experimentally to be negative.

So we find about $20\%$ deviation from the chiral normalization with
the $\rho$ resonance only, but $8\%$ deviation if we include
both the $\rho$ and the $\rho'$.
(Note that from the above, $\beta \approx - 0.10.$, where $\beta$ is
the quantity defined below in (\ref{eqnbeta}).)

There are three possible explanations for the remaining $8\%$ discrepancy.
\begin{enumerate}
\item there might be a small energy dependence of the meson couplings, 
i.e.\ the idea of chirally normalizing might be wrong.
\item it might be due to the $\rho''$, which we neglected.
\item there might be a non-resonant contribution, i.e.\ no full VMD.
\end{enumerate}

So we suggest three new parametrizations of the vector form factor
$F_V(Q^2)$. The original CN-VMD parameterization, as built into 
TAUOLA \cite{tauola}, is
\begin{equation} \label{eqnbeta}
   F_V^{(I)}(Q^2) = \frac{1}{1+\beta} \left[
   \BW_\rho(Q^2) + \beta \BW_{\rho'}(Q^2) \right]
\end{equation}
with
\begin{eqnarray}
   \beta = -0.145
\end{eqnarray}
In the second parameterization, we allow for a small energy dependence
of the meson couplings.
\begin{equation}
   F_V^{(II)}(Q^2) = \frac{1}{1+\beta} 
   \left[ \Phi(Q^2) \right]^2
   \left[
   \BW_\rho(Q^2) + \beta \BW_{\rho'}(Q^2) \right]
\end{equation}
where
\begin{eqnarray}
   \Phi(Q^2) & = & 1 + \frac{Q^2}{M^2}
\nonumber \\
   \beta & = & -0.07841
\nonumber \\
   M & = & 3.390 \unit{GeV}
\end{eqnarray}

Let us explain this parameterization. We assume that the meson couplings
are energy dependent
\begin{eqnarray}
   f_\rho & = & f_\rho(Q^2) = \Phi_{f}(Q^2) f_\rho(Q^2 = 0)
\nonumber \\
   g_{\rho\pi\pi} & = & g_{\rho\pi\pi}(Q^2) = \Phi_{g}(Q^2) 
   g_{\rho\pi\pi}(Q^2 = 0) 
\end{eqnarray}
and so on.

Now if the momentum dependence from $Q^2 = 0$ up to the resonance masses
is small, we can use a Taylor expansion
\begin{equation}
   \Phi_{X}(Q^2) = 1 + \frac{Q^2}{M^2} + \cdots
\end{equation}
Also, we make the simplifying assumption that the energy dependence
of all meson couplings is described by the {same} function $\Phi(Q^2)$.
With these assumptions, we can determine $\beta$ and $M$ by requiring 
$F_V$ to have the correct value at $Q^2 = 0$, and to be consistent 
with the on-shell values for the meson couplings.

Our second new parameterization has constant couplings, but not a full
VMD:
\begin{equation}
   F_V^{(III)}(Q^2) = 
   \alpha \BW_\rho(Q^2) + \beta \BW_{\rho'}(Q^2) + (1-\alpha - \beta)
\end{equation}
where
\begin{eqnarray}
   \alpha & = & 1.197
\nonumber \\
   \beta & = & - 0.116
\nonumber \\
   \Rightarrow 1 - \alpha - \beta & = & - 0.081
\end{eqnarray}

The third new parameterization includes the $\rho(1700) = \rho''$
\begin{equation}
   F_V^{(IV)}(Q^2) = 
   \alpha \BW_\rho(Q^2) + \beta \BW_{\rho'}(Q^2) + (1-\alpha - \beta)
   \BW_{\rho''}(Q^2)
\end{equation}
where
\begin{eqnarray}
   \alpha & = & 1.197
\nonumber \\
   \beta & = & - 0.116
\nonumber \\
   \Rightarrow 1 - \alpha - \beta & = & - 0.081
\end{eqnarray}

For practical purposes, it is important to note that the 
parametrizations (II), (III), and (IV) are numerically almost 
identical. This is to be expected, because they all have the same
limit at $Q^2 = 0$, and they are all consistent with the experimental
on-shell couplings.

\section{Comment on $\tau\to 4\pi \nu_\tau$}

From pQCD,
\begin{equation}
   \B(J^P = 1^-, S=0) = (31.9 \pm 0.1)\%
\end{equation}
The main final states which contribute to the non-strange vector channel 
are $2 \pi$, $4 \pi$, $K K$ and the vector part of $K K \pi$.
If we subtract the $2\pi$ and $KK$ contributions, we find
\begin{equation}
   \B(J^P = 1^-, S=0) - \B(2\pi) - \B(2K) = (5.28 \pm 0.36)\%
\end{equation}
This has to be saturated by $4\pi$ and $(K K \pi)_V$, any other 
final state will have negligibly small contribution.

$\B(K K \pi)_V$ is certainly not bigger than $0.22\%$. This is the
value from our paper \cite{kaon}, which definitely has to large
predictions for $K K \pi$ \cite{estes}. The experimental value
for the $4\pi$ contribution is
\begin{equation}
   \B^{RPP} (4\pi) = 5.39 \%
\end{equation}
In our paper on four pions \cite{fourpions}, however, 
we predicted 
\begin{equation}
   \B^{RPP} (4\pi) = 4.09 \%
\end{equation}
on the basis of new data for $e^+ e^- \to 4\pi$ from Orsay.

We can now see that pQCD requires $\B(4\pi)$ to be about $5\%$,
as the RPP-96 says, and not about $4\%$, as we predicted in
\cite{fourpions}.

\section{Summary and Conclusions}
%
We discussed and tested the assumptions of the chirally normalized
vector meson dominance model (CN-VMD).

To test whether the meson couplings are really constant, we
compared the CN-VMD amplitudes with those derived using on-shell
couplings directly from experimental data. We discuss in detail the 
$3\pi$, $K K \pi$ and $2\pi$ final states. 
For the case of the $3\pi$, we found reasonable agreement. 
The discrepancy was much larger in the case of the $K K \pi$ mode.
We suggested that this might actually be due to the possibility
that the 
true values for the $K_1$ widths might be larger than those
quoted in the Review of Particle Physics.
We also proposed new parametrizations for the vector form factor
in the $2\pi$ mode, and we confirmed that the branching ratio
into four pions should be indeed around $5 \%$.

\section*{Acknowledgments}
%
This work is supported in part by the National Science
Foundation (Grant \#PHY-9218167), and by the
Deutsche Forschungsgemeinschaft.
The author would like to thank the members of the 
Theoretical Physics Group at Harvard
University for their kind hospitality.

\appendix
\section{Appendix}
%
\subsection{Feyman-Rules}

We define the various couplings 
by the following Feynman rules:
\begin{equation}
\begin{array}{ccc}
   \tau \to \nu_\tau \rho & & i \frac{G_F}{\sqrt{2}} f_{\rho} \cos \theta
   [\gamma_\mu \gamma_- ]
\\ \\
   \tau \to \nu_\tau a_1 & & i \frac{G_F}{\sqrt{2}} f_{a_1} \cos \theta
   [\gamma_\mu \gamma_- ]
\\ \\
   \tau \to \nu_\tau K_1 & & i \frac{G_F}{\sqrt{2}} f_{K_1} \sin \theta
   [\gamma_\mu \gamma_- ]
\\ \\
   (a_1^-)_\mu \to (\rho^0)_\nu \pi^- & &
   i g_{a_1 \rho \pi} g_{\mu \nu}
\\ \\
   (K_1^-)_\mu \to (K^{*0})_\nu \pi^- & &
   i g_{K_1 K^* \pi} g_{\mu \nu}
\\ \\
   (K_1^-)_\mu \to (\rho^0)_\nu K^- & &
   i g_{K_1 \rho K} g_{\mu \nu}
\\ \\
   (\rho^0)_\mu \to \pi^+(p_1) \pi^-(p_2) & &
   i g_{\rho\pi\pi} (p_1 - p_2)_\mu
\\ \\
   (\overline{K^{\star 0}})_\mu \to \pi^+(p_1) K^-(p_2) & &
   - i g_{K^* K \pi} (p_1 - p_2)_\mu
\\ \\
   (\overline{K^{\star 0}})_\mu \to \pi^0(p_1) \overline{K^0}(p_2) & &
   - i \frac{g_{K^* K \pi}}{\sqrt{2}} (p_1 - p_2)_\mu
\end{array}
\end{equation}

\subsection{Meson-Gauge Boson Couplings}
\begin{eqnarray}
   f_{\pi} & = & (92.1 \pm 0.3) \unit{MeV} 
\\[1ex]
   f_K     & = & (112.4 \pm 0.9) \unit{MeV}
\\[1ex]
   f_{\rho} & = & (0.1167 \pm 0.0025) \unit{GeV}^2
\\[1ex]
   f_{\rho'} & = & \left\{
   \begin{array}{ll}
   (0.184 \pm 0.007) \unit{GeV}^2 & \mbox{from $e^+ e^- \to 2 \pi$} \\
   (0.1675 \pm 0.0010) \unit{GeV}^2 & \mbox{from Weinberg's sum rules} \\
   (0.1758 \pm 0.0082) \unit{GeV}^2 & \mbox{average of the two} \\
   \end{array} \right.	
\\[1ex]
   f_{a_1} & = & 0.2050 \unit{GeV}^2
\\[1ex]
   f_{K_1} & = & (0.2026 \pm 0.0103) \unit{GeV}^2
\end{eqnarray}

\subsection{Triple Meson Couplings}
\begin{eqnarray}
  g_{\rho\pi\pi} & = & 6.049 = \sqrt{2} g_{VPP}, \quad 
  g_{VPP} = 4.277
\\[1ex]
  g_{\rho' \pi \pi} & = & 1.39
\\[1ex]  
  g_{a_1\rho\pi} & = & (4.33 \cdots 5.30) \unit{GeV}
  \quad \mbox{for $\Gamma_{a_1} = (400 \cdots 600) \unit{MeV}$}
\\[1ex]
  g_{K^*K\pi} & = & 4.578 = (1 + \epsilon) g_{VPP}, \quad
  \epsilon = 0.07
\\[1ex]
  g_{K_1 K^* \pi} & = & 3.535 \unit{GeV} \quad \mbox{if $\Gamma_{K_1} = 
  174\unit{MeV}$}
\\[1ex]
  g_{K_1 \rho K} & = & 1.5848 \unit{GeV} \quad \mbox{if $\Gamma_{K_1} =
  90 \unit{MeV}$}
\end{eqnarray}


\end{document}